\documentclass[12pt]{elsarticle}

\usepackage{graphicx}
\usepackage{amssymb}
\usepackage[T1]{fontenc}
\usepackage[utf8]{inputenc}
\usepackage{amsfonts}
\usepackage{amssymb}
\usepackage{amsthm}
\usepackage[centertags]{amsmath}
\usepackage{latexsym}								
\usepackage{lettrine}
\usepackage{graphicx}
\usepackage{subcaption}
\usepackage{adjustbox}
\usepackage{fancyhdr}

\usepackage{afterpage}

\usepackage{tabularx}
\usepackage{color}

\usepackage[table]{xcolor}
\definecolor{KUBlue}{RGB}{0,34,180}
 \definecolor{KUCrimson}{RGB}{232,0,13}

\usepackage{geometry}
\usepackage{pdflscape}

\usepackage[figuresright]{rotating}	

\usepackage[section]{placeins}			
\usepackage{array}									
\usepackage{lscape}									
\usepackage{multirow}								
\usepackage{makeidx}								
\usepackage{mdwlist}
\usepackage{longtable}
\usepackage{supertabular}						
\usepackage{booktabs}
\usepackage{tabularx}
\usepackage{microtype}
\usepackage{relsize}
\usepackage{csquotes}

\usepackage[acronym]{glossaries}

\usepackage{tabularx,ragged2e,booktabs}

\newcolumntype{L}{>{\RaggedRight\arraybackslash}X} 

\usepackage{rotating}
\usepackage[breaklinks,colorlinks=true,linkcolor=black,citecolor=black]{hyperref}
\usepackage[pagewise,switch]{lineno}	 	

\usepackage{enumerate}							
\usepackage{chemformula}

\usepackage{appendix}
\usepackage{blindtext}

\usepackage{lineno}

\journal{None}

\begin{document}

\begin{frontmatter}


\title{Mobile Phone Metadata for Development}



\author{Damien C. Jacques}

\address{Louvain-la-Neuve, Belgium}

\begin{abstract}

Mobile phones are now widely adopted by most of the world population. Each time a call is made (or a SMS sent), a Call Detail Record (CDR) is generated by the telecom companies for billing purpose. These metadata provide information on when, how, from where and with whom we communicate. Conceptually, they can be described as a geospatial, dynamic, weighted and directed network. Applications of CDRs for development are numerous. They have been used to model the spread of infectious diseases, study road traffic, support electrification planning strategies or map socio-economic level of population. While massive, CDRs are not statistically representative of the whole population due to several sources of bias (market, usage, spatial and temporal resolution). Furthermore, mobile phone metadata are held by telecom companies. Consequently, their access is not necessarily straightforward and can seriously hamper any operational application. Finally, a trade-off exists between privacy and utility when using sensitive data like CDRs. New initiatives such as Open Algorithm might help to deal with these fundamental questions by allowing researchers to run algorithms on the data that remain safely stored behind the firewall of the providers.
\end{abstract}


\end{frontmatter}



\section{Introduction}

Over the past two decades, access to telecommunication services has seen exponential growth. From around 100 million in 1995, the number of mobile cellular subscriptions has risen to 7.4 billion  worldwide in 2016 -- the equivalent of the entire world population \citep{ITU2016}. No technology has ever spread faster around the world \citep{economist2008}. This growth was primarily driven by wireless technologies and liberalization of telecommunication markets, along with new financing and technology, which have enabled faster and less costly network rollout. \\

However, this does not mean that every person in the world has subscribed to a mobile service. Because many individuals own several handsets or have multiple subscriber identity module (SIM) cards, the number of subscribers, estimated to 4.7 billion worldwide, is substantially lower than the number of subscriptions \citep{intelligence2016mobile}. This is because the number of subscriptions tend to exaggerate the mobile phone penetration rate\footnote{The mobile phone penetration rate is the number of active mobile phone users per 100 people within a specific population.} in developed economies. On the other hand, in many developing countries, mobile phone access is higher than subscription numbers would suggest. Access is indeed fostered in countries where sharing mobile phones is a common practice, especially within large households \citep{aker2010mobile}. In a world bank report, the practical impact of the difference between subscription and household penetration\footnote{Portion of total households having access to mobile phone within a specific population.} is clearly explained \citep{world2012information}: 

\begin{displayquote}
“Take Senegal, where the subscription penetration was 57 per 100 people in 2009, but household penetration was estimated to be 30 points higher at 87. This larger household size can dramatically extend access to mobile phones, considering that on average nine persons are in each Senegalese household.“ It results that “several low-income nations have higher mobile phone home penetration than some developed economies. For example, Senegal, along with some other low- and middle-income economies, has a higher proportion of homes with mobile phones than either Canada or the United States“.
\end{displayquote}
 
The economic potential of the mobile phone is tremendous. 2015 has been a year of continued growth in the mobile industry, with operator revenues exceeding \$1 trillion.  The mobile ecosystem generated 4.2\% of Gross World Product and directly support 17 million jobs \citep{intelligence2016mobile}. The mobile telephony is increasingly recognized as an essential tool of development by improving the flow of information and providing a platform for financial services \citep{aker2014economic}. The more striking example lies in the mobile money service which is now widely established and brings financial inclusion to previously unbanked and underbanked populations across the developing world (1.9 billion people globally). Mobile phones are also a key platform to bring internet access to people across the globe, particularly in developing regions where fixed broadband services are prohibitively expensive, and fixed-line infrastructure is limited.  At the end of 2015, 2.5 billion individuals from the developing world were accessing the internet through mobile devices \citep{intelligence2016mobile}. \\

Modern mobile phones (smartphones) are now integrated computers with dozens of embedded sensors, such as accelerometer, digital compass, gyroscope, GPS, microphone, and camera, which enable the emergence of several research applications based on personal sensing \citep{lane2010survey}. These are promising avenues that are believed to revolutionize many sectors of the economy, but the penetration of smartphones is still low in Africa because of their higher cost. For example, only 19\% of the Senegalese population reported owning a smartphone in 2015 \citep{poushter2016smartphone}. \\

On the other hand, even the most basic handset passively generates a vast amount of metadata leaving behind a digital trace\footnote{Also called digital shadow, digital footprint or data exhaust.} of the activity of its user. These metadata provide information on when, how, from where and with whom we communicate \citep{blondel2015survey}. At first, researchers realized the potential of such data by uploading tracking software into consenting subjects’ phones through the Reality Mining project of the MIT\footnote{Note that as early as 1999, it was already demonstrated that OD matrix could be obtained from the localisation of mobile phones \citep{white2002extracting}. But the potential of CDRs for computational social science was discovered later.} \citep{eagle2006reality}. They later gained access to actual metadata directly from mobile network providers, leading to larger-scale research and greater analytical power (e.g., \citealp{gonzalez2008understanding}). Until then, more and more datasets were opened up to the scientific community, and mobile phone metadata are now seen as a typical example of empirical data used in network science \citep{barabasi2016network}. The applications are tremendous, particularly for studies related to mobility, social network and socio-demographics of people \citep{blondel2015survey, saramaki2015seconds, naboulsi2016large}. Several initiatives have emerged, such as the Data For Development (D4D) challenge organized by Orange, that provided datasets to the research community for projects related to development. In a recent survey carried out by the World Bank, mobile phone data appeared at the top position, just before satellite imagery, in the Big Data sources used in SDG-related projects \citep{Ballivian2014}. \\

The objective of this paper is to introduce mobile phone metadata, in particular call data records used by the companies for billing purpose, and their potential for the SDGs. We first present some elements of mobile network infrastructure, a prerequisite to understand the characteristics of data collected by a mobile network operator. We then describe the specific features that make CDRs unique and how they can be used to help achieving the SDGs. Finally, we discuss the statistical limitations of such data and the risks associated with their use (in particular, data access and privacy).






\section{Elements of Mobile Network Operator Infrastructure}

This section has been written based on material developed in \citet{swenson2006gsm, tiru2014overview,ricciato2015estimating,janecek2015cellular,ricciato2017beyond}. \\

Almost all Mobile Network Operators (MNO) in the world use two main mobile technologies – GSM (Global System for Mobiles) and CDMA (Code Division Multiple Access)\footnote{CDMA may disappear in favor of GSM network due to the spread of the fastest and high-quality Long Term Evolution (LTE) technology that uses a similar technology as GSM.}. The market share of subscribers using CDMA worldwide is 15-25\% (mostly in North-America and some Asian countries) and 75-85\% for GSM (the rest of the world). The main difference between these technologies is the radio signaling technology. The practical implication is that a mobile device is tied to a particular network within CDMA network, while a Subscriber Identity Module (SIM) card is tied to a specific network within GSM network. It is, therefore, easier to switch mobile devices within GSM networks thanks to SIM card's portability. GSM phones without CDMA support cannot run within CDMA network (and conversely). \\

\begin{figure*}[t!]
    \centering        
        \includegraphics[width=1\linewidth]{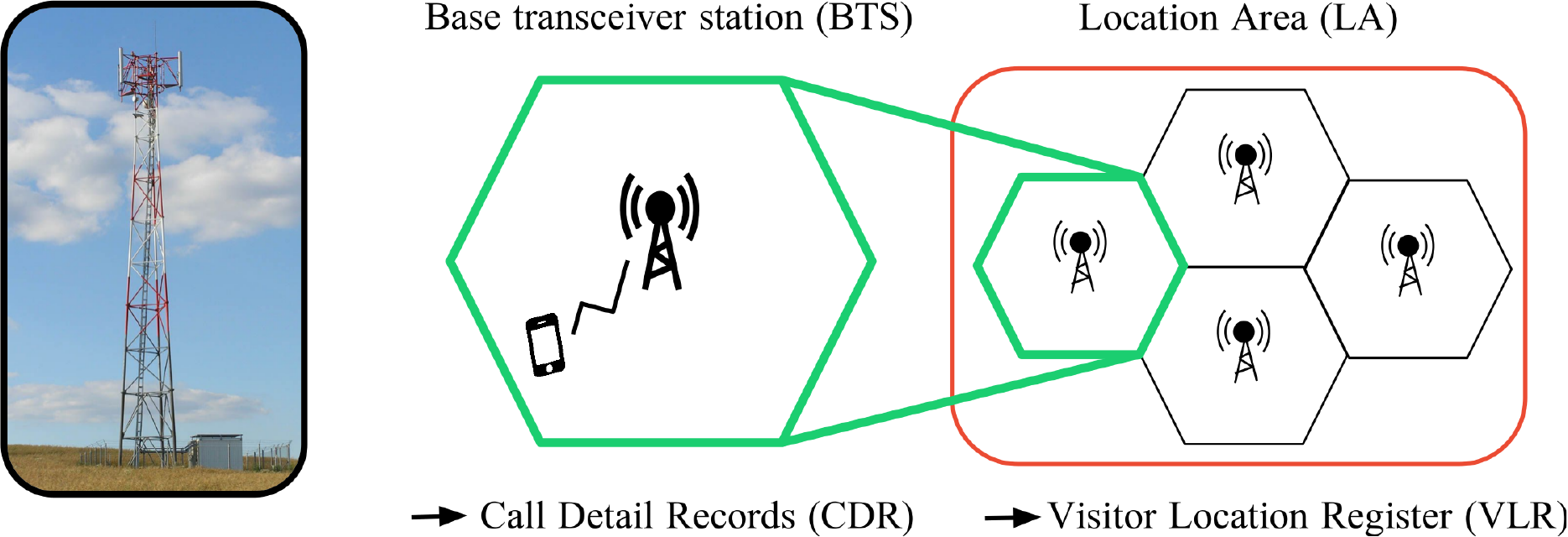} \\ 
    \caption{The two main sources of location data collected by Mobile Network Operators: the Base Transceiver Station (shown in green and picture on the left) is stored in the Call Detail Records and the Location Area (shown in red) is stored in the Visitor Location Register (VLR).}
    \label{fig:antennaANDla}
\end{figure*}

\begin{sidewaysfigure}
    \centering        
        \includegraphics[width=0.85\linewidth]{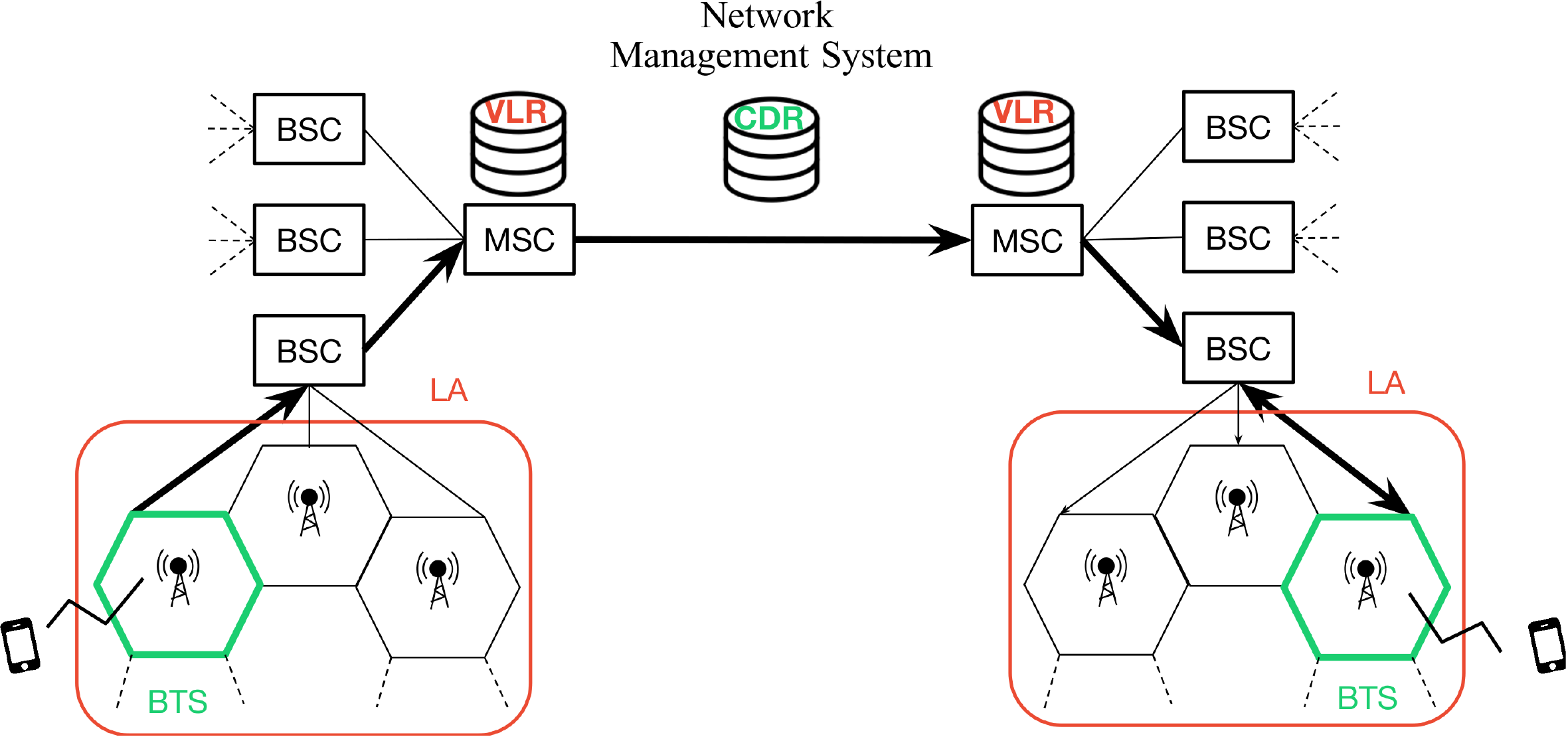} \\ 
    \caption{Simplified structure of a GSM network. MSC stands for Mobile Switching Center, BSC for Base Station Controller, LA for Location Area, BTS for Base Transceiver Station, VLR for Visitor Location Register and CDR for  Call Detail Records. Arrows indicate the propagation of the signal needed to locate the recipient when a call is initiated. }
    \label{fig:call}
\end{sidewaysfigure}

A GSM network is a radio network of individual cells, known as Base Transceiver Station (BTS). The BTS is responsible for transmitting and emitting radio communications between the network and the mobile devices which on the ground can be identified by the antenna tower and equipment (Figure \ref{fig:antennaANDla}, left). To improve the network efficiency, BTSs are hierarchically grouped in Location Area (LA) and controlled by a Base Station Controller (BSC) (Figure \ref{fig:antennaANDla} and \ref{fig:call}). The BSC is responsible for handover procedures\footnote{i.e. BTS switch when a call in progress moves from one base station to another.} within a single LA between one BTS to another. BSCs are controlled by Mobile Switching Centres (MSCs) that accommodates the Visitor Location Register (VLR) – the registry for holding the information about the LA in which the mobile devices is located (Figure \ref{fig:call}). Finally, MSCs report to the Network Management System (NMS) where all administrative and central procedures reside. Usually, only data that are transmitted from MSCs to NMS are stored for different purposes while lower level data traffic are deleted. NMS accommodates different registries and databases that are important for the network functioning, in particular, billing databases (Call Data Records\footnote{Call Detail Records and Call Data Records are used interchangeably in this paper.}). Components of Mobile Positioning System, the system used for pinpointing users’ location for emergency and security services (e.g., E112 and E911 directives in EU and US), also resides in NMS and MSCs. \\

Different MNOs may share some of the network equipment. For example, it is not uncommon that different MNOs share the same BTS. There are also special types of MNOs (virtual MNOs) that do not possess any network infrastructure, but instead rent it from other MNOs. In such case usually the virtual MNOs do not have access to all operational data.\\

Mobile phones that are switched on are either in idle or in active state. In idle state a mobile phone is not allocated any radio resource, but it constantly evaluate if it needs to switch to another cell with a better signal strength (it listens but does not transmit). Thus, in idle state a mobile phone receives passively, which implies that the network is unable to identify cell changes of idle mobile phones, except when such a switch is explicitly requested by the mobile phone. Cell switches within one LA (from one BTS to another) are not reported, but cell shifts from one LA to another are. It turns out that the LA of any switched-on mobile phone is known at any time by the MNO.\\

A mobile phone remains most of the time in idle state and only become active during a call or a data transaction. When a mobile phone user dials a number to make a call, a call initiation request is sent to the MSC. The MSC validates the request by checking the user's identity and airtime balance in the records of its database. If valid, a connexion is initiated with the third party, and the MSC requests the base station to move the mobile phone to an unused voice channel so that the call can begin. Once a call is in progress, the MSC adjusts the power transmitted by the mobile phone as it moves in and out of the coverage area of each base station. When a mobile phone with a call in progress moves from one BTS to another, handover procedure are automatically managed by the network. \\

In summary, the state of the mobile phone (idle or active) determines the temporal and spatial accuracy of the user location within the network data system.\\

\section{Mobile Phone Metadata}





\subsection{Call Detail Records (CDR)} 

CDRs typically include BTS location information of the caller and recipient (starting cell), as well as time stamp and call duration (Table \ref{tab:Sample_CDR}). However, the information stored in CDRs really depends on each MNO as no standardised structure exists (e.g., whether long calls are chunked into multiple CDRs) \citep{tartarelli2010lessons}. Contrary to what the term suggests, CDRs are also used for SMS and data connection (sometimes stored in separate data files). CDR data contain cell-level locations, but only for \textit{active} mobile phone engaged in voice call, SMS or data connections. Their use for billing purpose implies an archived and a constant update of the data without changing nor deleting old records. This means that long time series are easily accessible for data analytics.  \\

\begin{table}[b!]
\centering
\footnotesize
\caption{Sample of typical call data records.}
\label{tab:Sample_CDR}
\begin{tabular}{cccccc}
\toprule
Caller &  Callee  & Outgoing & Incoming & Timestamp           & Call duration  \\
SIM  & SIM  &  BTS  &  BTS  &            & (sec) \\
\midrule
0458685984       & 0488595496       & 12                  & 365                 & 2018-01-18 15:22:12 & 456                 \\
0458685984       & 0458685984       & 12                  & 25                  & 2018-01-18 22:24:12 & 35                  \\
0469875254       & 0498563201       & 879                 & 567                 & 2018-01-19 08:47:10 & 125                 \\
(...)            & (...)            & (...)               & (...)               & (...)               & (...)              \\
\bottomrule
\end{tabular}
\end{table}


\subsection{Visitor Location Register (VLR)} 

The VLR is a dynamic database supporting the operation of the Mobile Switching Center (MSC). Principally, the VLR caches temporary data about the current LA location of all mobiles, both active and idle.  Due to the completeness of VLR data, VLR records provide an instantaneous description of the location of all mobile phones, at LA level. VLR data are highly dynamic because LA locations are updated continuously. As a result, bulk VLR reading can only be conducted in real-time in the background of network operations. \\

\subsection{Passive Monitoring Systems} 

Some MNOs track signaling and traffic exchange (e.g. for handover) in the network through passive monitoring systems. These systems aim to assess and resolve network operation and troubleshooting \citep{tartarelli2010lessons}. Using the network data, passive monitoring systems are able to locate every mobile phone with great accuracy, both in terms of time and
space. Locations at cell and LA level are provided, both for active and for idle mobile phones. \\

\subsection{Selecting a dataset} 

Selecting a data source for analysis is a trade-off between spatial and temporal resolution, and data accessibility \citep{ricciato2017beyond}. CDRs provide information at the highest spatial resolution (cell level, see Figure \ref{fig:location_precision}) but are event-based. Therefore, data are only available when the user makes a call (or send an SMS/data). This can be a limitation for mobility studies that require regular location update (see Statistical limitations section). On the other hand, the mobile phone usage (number of calls, when and where, etc.) might provide precious information on the user's socio-demographic profile. CDRs are stored offline and therefore easily accessible. Because of that, these are, by far, the data most frequently used in the literature \citep{blondel2015survey}. \\

VLR provide data at the finest temporal resolution but at LA level. The spatial resolution of LA is much lower than BTS. For instance, the Ile-de-France region has almost 10,000 BTS grouped in only 32 LAs, each of which has between 150 and 500 BTS \citep{bonnel2015passive}. \\

Lastly, passive monitoring systems leveraging signaling data combine the best of the two worlds. However, the systematic acquisition of such data are generally based on proprietary system and not available to all operators. As a result, very little research has been conducted using these data \citep{janecek2015cellular, valerio2009exploiting}.\\

Other data collected by MNO that are relevant for research applications include \textbf{airtime credit purchases} and \textbf{customer client profiles}. Airtime credit purchases are useful to predict socio-economic status of population \citep{gutierrez2013evaluating, decuyper2014estimating, blumenstock2015calling} while customer client profile enrich mobile phone metadata with personnal information such as gender and age. However, while mobile phone operators have access to all the information filed by their customers and their operational data, they generally limit the access to only a sample of what is available depending on their own privacy policies and the regulation on privacy protection of each country. \\

The following sections further detail the specific characteristics of CDRs.

\section{Data Features}

Conceptually, CDRs can be described as a geospatial, dynamic, weighted and directed network (Figure \ref{fig:networkCDR} -- A). In the following sections, each of these dimensions is further developed.

\subsection{Network Dimension}

\begin{figure*}[t!]
    \centering        
        \includegraphics[width=0.8\linewidth]{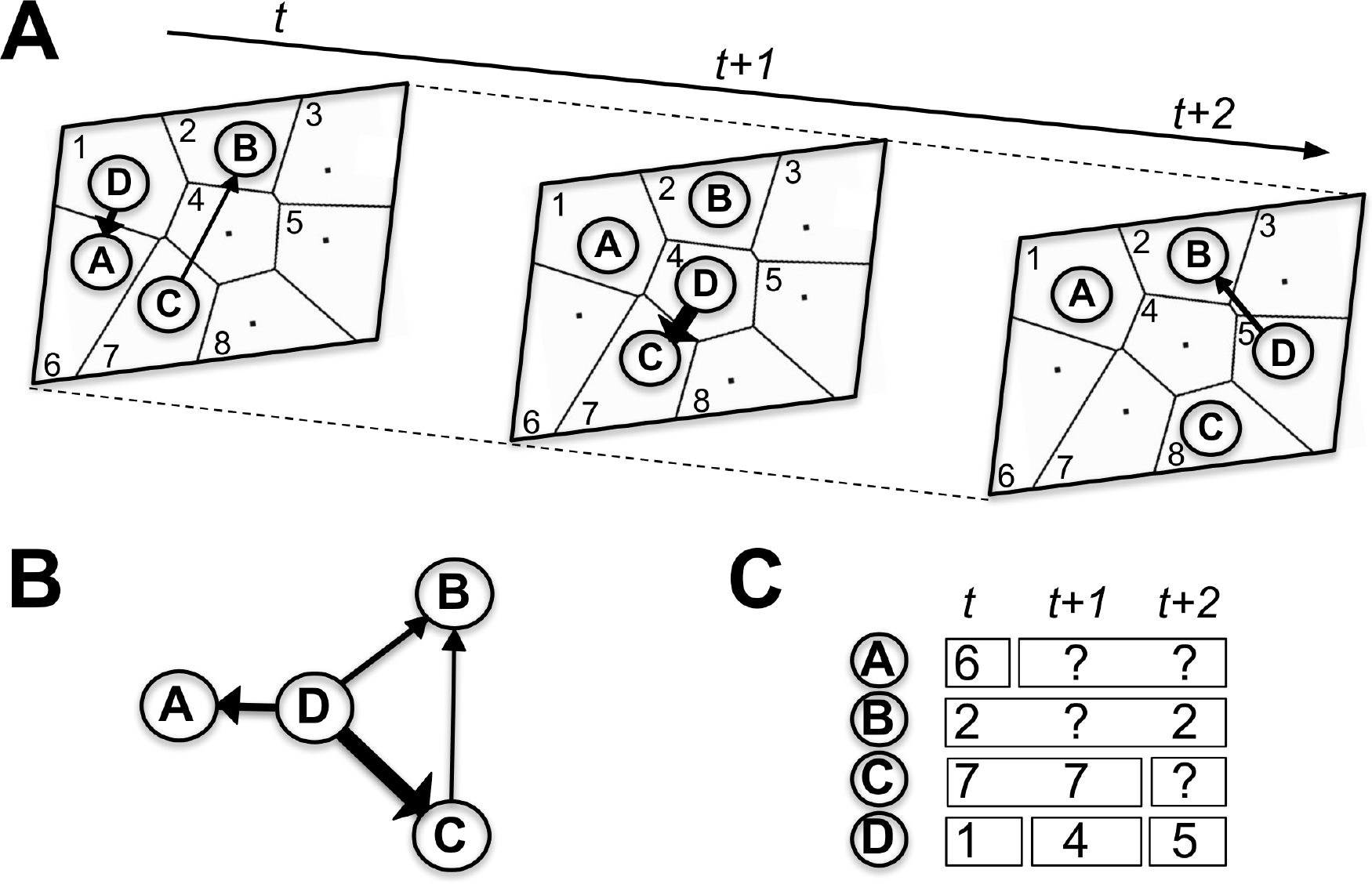} \\ 
    \caption{Schematic representations of CDRs data. Letters (A-D) represent SIM cards ($\sim$ individuals), numbers (1-8) represent antenna coverage approximated by a Voronoï tesselation, and arrows represent call direction (head) and duration (width). (A) geo-spatial, dynamic, directed weighted network (here weights are call duration), (B) static, directed weighted network (over $t$ to $t+2$ period), and (C) dynamic trajectories of SIM cards.}
    \label{fig:networkCDR}
\end{figure*}

A network is defined by its nodes (or vertex) and links (or edges, ties). In CDRs, nodes are SIM cards ($\sim$ mobile phones) and links are the calls (or SMS) exchanged between two SIMs. Furthermore, the links are directed from the caller to the callee and the duration of calls weights each tie (Figure \ref{fig:networkCDR} -- A). By neglecting the geospatial dimension and aggregating (e.g., by taking the sum) the total call duration between each pair of users over a given period, a static representation of the CDRs network can be obtained (Figure \ref{fig:networkCDR} -- B). After temporal aggregation, the link weights of the network might also represent the number of calls or SMS exchanged between two users. \\

According to graph theory, such network can be represented through its adjacency matrix $A_{ij}$ (Eq. \ref{eq:adjacency}). 

\begin{equation}
A_{ij} = 
 \begin{pmatrix}
  0 & A_{1,2} & \cdots & A_{1,n} \\
  A_{2,1} & 0 & \cdots & A_{2,n} \\
  \vdots  & \vdots  & \ddots & \vdots  \\
  A_{m,1} & A_{m,2} & \cdots & 0 
 \end{pmatrix}
 \label{eq:adjacency}
\end{equation}

The values of $A_{ij}$ represent the weight of the links (call duration, number of calls or number of SMS) between node $i$ and $j$. Because the network is directed, the matrix is asymmetric and $A_{ij}\neq A_{ji}$. Note that $A_{ii}=0$ because one user cannot call himself. \\

The topological structure (degree, density, connectivity, etc.) of CDRs network inform on the characteristics of people's social networks. Such analysis lead to the emergence of a new field of research called computational social science \citep{lazer2009life}. One interesting possibility offered by CDR network is the detection of social communities by identifying group of users that interact more with each other than with the rest of the population. This requires efficient algorithm capable to handle large dataset (see for instance the 'Louvain method' in \citealp{blondel2008fast}).

\subsection{Geospatial Dimension}

\begin{figure*}[b!]
    \centering        
        \includegraphics[width=1\linewidth]{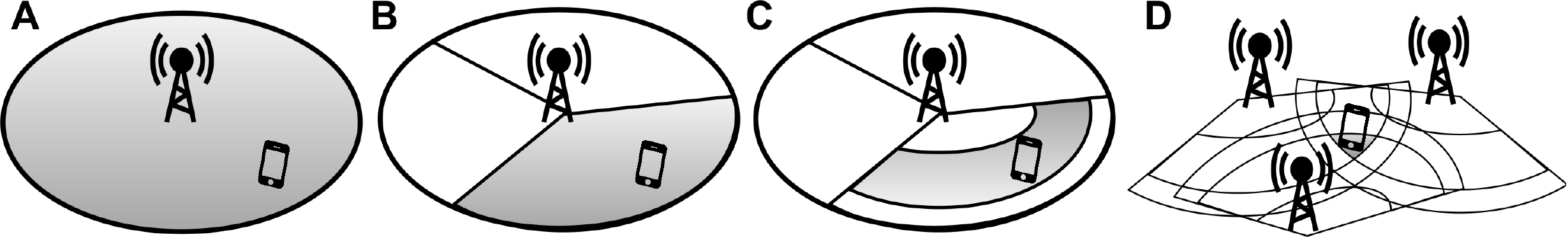} \\ 
    \caption{Location at (A) the base station level, (B) the sector level, (C) the sector level knowing signal characteristics, (D) triangulation.}
    \label{fig:location_precision}
\end{figure*}

\begin{figure*}[t!]
    \centering      
        \includegraphics[width=0.8\linewidth]{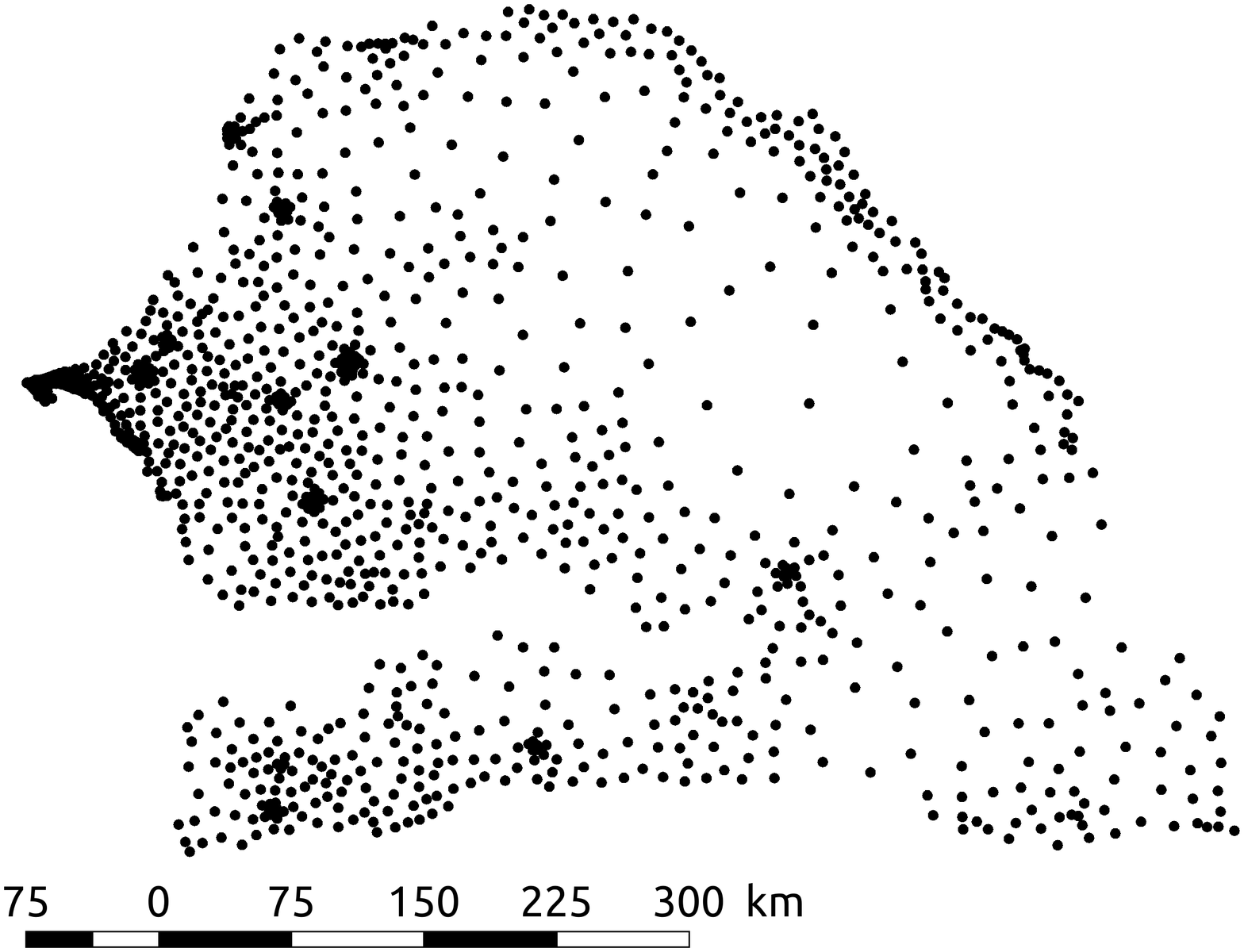} \\
        \includegraphics[width=0.8\linewidth]{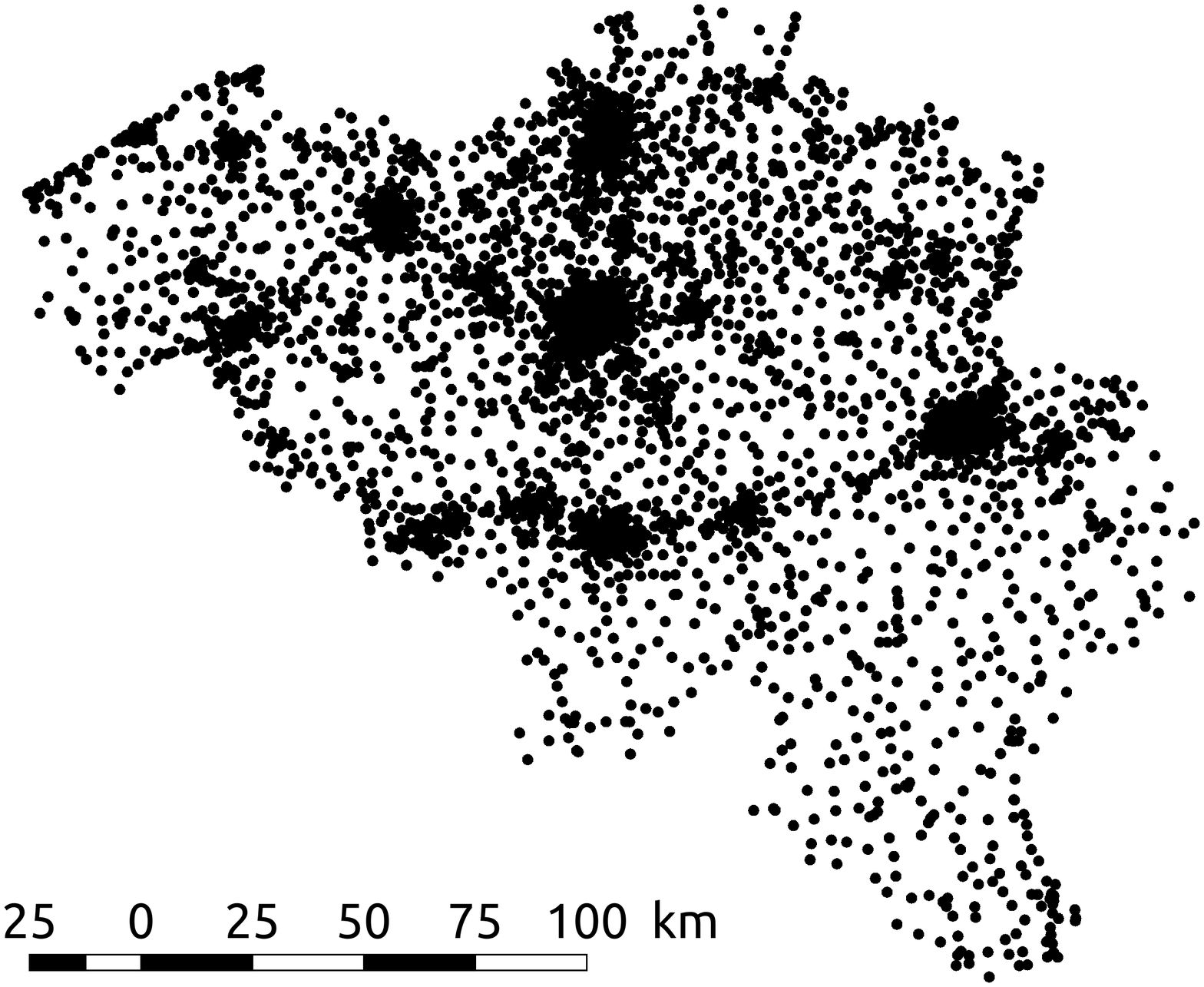} \\
    \caption{BTS maps of Orange Sonatel in Senegal (top) and Orange Mobistar in Belgium (bottom).}
    \label{fig:mapAntennna}
\end{figure*}

As previously explained, each call in CDRs is geolocated at the BTS level (Figure \ref{fig:location_precision} -- A) so that a mobile phone user can be located within the coverage of this BTS. One BTS generally accommodate more than one antenna (typically three). Therefore, the location of the user can be more precisely defined within the coverage of one antenna (Figure \ref{fig:location_precision} -- B). However, MNOs might only provide CDRs at BTS scale to preserve anonymity (e.g., the D4D dataset). The coverage of one particular antenna depends on its technical characteristics (power, technology, etc.) and the BTS density (Figure \ref{fig:mapAntennna}). More antennas are required in areas with more traffic which means that in urban and sub-urban areas, cell areas typically span between hundreds of meters (micro-cells) and a few kilometers of diameter, while sparsely populated areas are covered by few macro-cells. Smaller cells (pico-cell and femto-cells) can also be deployed in highly crowded areas, such as shopping malls, train stations, or airports. The antenna density is still considerably lower in developing countries than in developed areas (see for instance Figure \ref{fig:mapAntennna}). It is worth mentioning that analyzing the characteristics of the signal exchanged between the phones and the BTS allow to reduce spatial uncertainty by inferring the user-antenna distance (using response-delay and strength) (Figure \ref{fig:location_precision} -- C). Finally, triangulation can further increase the spatial accuracy of the estimation of the user's location (Figure \ref{fig:location_precision} -- D). Such procedure usually requires authorization and approval from the user except for emergency response such as E112 and E911 directives in EU and US.

\subsection{Temporal Dimension}

The temporal resolution of CDRs depends on the activity of mobile phone users. The best case is a user multiplying short event (SMS or short calls). It is worth noting that a long call does not increase the temporal resolution as most of the time, only the starting cell (where the call was initiated) is recorded in CDRs. While it is a limitation for mobility analysis, the temporal patterns of mobile phone usage is interesting to study the socio-economic profiles of population \citep{blumenstock2015predicting,Soto:2011}. \\

\begin{figure*}[t!]
    \centering        
        \includegraphics[width=0.8\linewidth]{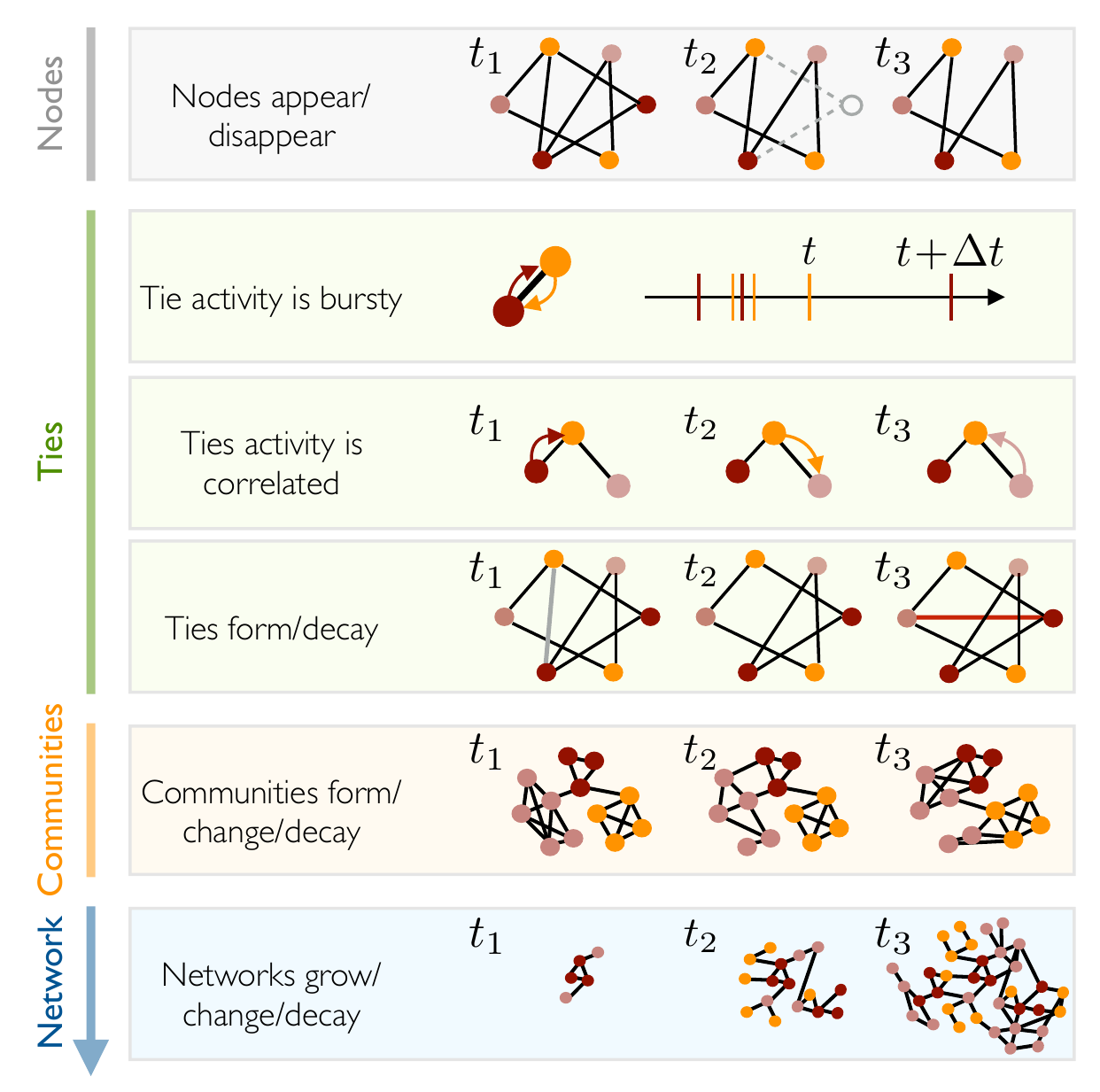} \\ 
    \caption{Temporal features of network at different structural and spatial scales. Figure from \citet{saramaki2015seconds} reproduced with permission of the authors.}
    \label{fig:timescales}
\end{figure*}

On the other hand, the temporal dimension of CDRs allows studying the dynamic of social networks. \citet{saramaki2015seconds} showed the great value of this approach for different scales of analysis (Figure \ref{fig:timescales}). For instance, one can explore the resilience of social ties for an individual or a community after a shock such as a loss (at individual level) or a disaster (at community level).\\

Finally, one key element of CDRs is their potential for near real time applications. As the data are collected on the fly, they are virtually instantly available for analysis. This aspect has significant implications for post-disaster monitoring and early warning systems.

\subsection{Spatio-Temporal Dimension}

The combination of both spatial and temporal dimension provide valuable information for dynamic population mapping and mobility analysis as well as for land use classification. \\

The most straightforward application of CDRs is the estimation of the number of people at specific place and time. This implies modeling the relationship between the number of active mobile phone users with the actual population over time and space. Using census as calibration data, accurate estimations of population can be obtained during night-time \citep{deville2014dynamic}. However, the fact that the relationship between active mobile phone users and the actual population is not independent of time and space makes it harder to accurately extrapolate these models during day-time. This would require calibration data for daytime population which are rarely available.\\

From the CDRs, a spatial trajectory can be computed for each user and use in mobility analyses (Figure \ref{fig:networkCDR} -- C). The quality of the inference of such an approach depends on the frequency of the user activity (e.g., in Figure \ref{fig:networkCDR} -- C, mobility of user D will be better predicted than user A). In particular, mobile phone activity is known to be bursty \citep{barabasi2005origin}. Users tend to place most of their calls in short bursts, followed by long periods with no call activity, during which information about the user’s location is lacking\footnote{This observation also applies to several other human activities such as web browsing, stock trading or library visits \citep{vazquez2006modeling, barabasi2010bursts}.}. However, despite their temporal sparseness and spatial coarseness, CDRs still offer great insights into the movement patterns of individuals and communities \citep{becker2013human}. Furthermore, \citet{song2010limits} showed that human mobility was highly predictable, regardless of the distance traveled, due to the regularity of our daily mobility. A typical application of mobility analysis with CDRs is the computation of origin-destination matrices at different temporal scales. It allows identifying different mobility dynamics such as daily commuting and long-term migration. \\

On the other hand, the temporal signature of BTS activity can be used to define land use patterns. This method has been applied in urban areas to make the distinction,  between residential, business, industrial and leisure areas among others \citep{lenormand2015comparing}.


\section{Applications of Mobile Phone Metadata for Development}



The use of CDRs for data for development is relatively new. The growing interest in the field has been triggered by telecom companies opening large datasets to the research community. The Data For Development (D4D) challenge, launched by Orange in 2013, was the first release of an extensive CDR database from an African country (Ivory Coast) to the international research community \citep{blondel2012data}. It was also the first CDRs-related project to be labeled as ‘development’, and gained huge publicity after the United Nations, the World Economic Forum and several high-profile academic institutions (including MIT and Cambridge University) endorsed it \citep{taylor2015}. The initiative was very successful and resulted in dozens of innovative projects developed by research labs from around the world. Encouraged by this success, a second D4D challenge, with data from Senegal, was organized in 2014. Once again, it resulted in several creative projects \citep{de2014d4d}. The pioneer work of the Big Data for Good team of Telefonica\footnote{Telefonica is a Spanish multinational broadband and telecommunications provider serving over 200 million users in Latin America, Europe, and the United States.} in the development of algorithms for inferring socio-economic welfare from mobile phone use patterns should also be acknowledged. \\

Hereafter, we briefly present five contrasted applications of CDRs analysis for development in five different developing countries (Kenya, Haiti, Rwanda, Ivory Coast, Senegal). The aim here is not to give an exhaustive review of all the possible applications of CDRs but rather a quick overview of what can been done. For a detailed and broad review of CDRs data analysis, the reader is referred to the works of \citet{blondel2015survey}, \citet{naboulsi2016large} and \citet{saramaki2015seconds}.\\

\subsection{Health}

One of the most common (and successful) use of CDRs data for development is for epidemiological studies of human infectious diseases. For instance, \citet{wesolowski2012quantifying} used CDRs from Kenya to identify the dynamics of human carriers that drive parasite importation between regions. They analyzed the regional travel patterns of nearly 15 million individuals over the course of a year and characterized the degree of connectivity among different areas in Kenya. Using a simple transmission model and malaria infection prevalence data, they were then able to map the importation routes that contribute the most to malaria epidemiology on regional spatial scales.

\subsection{Post-Disaster Management} 

Another application that greatly benefits from mobility data derived from CDRs is crisis management following a disaster.  \citet{lu2012predictability} analyzed the movements of nearly two million SIM card holders before and after the 2010 earthquake in Haiti, finding that one-fifth of Port-au-Prince’s residents left the city by three weeks after the disaster. They also show that the trajectory of people fleeing from regions hit by the earthquake was highly correlated to their mobility patterns during normal times. Such findings suggested that population movements during disasters are significantly more predictable than previously thought and highly influenced by people’s social support structures.

\subsection{Poverty and Socio-Economics Level}

Assessing socio-economics levels, in particular poverty prevalence, is another recent development of mobile phone metadata. \citet{blumenstock2015predicting} showed how the individual’s past history of mobile phone use can inferred his/her socio-economic status using records of billions of interactions on Rwanda’s largest mobile phone network. They validated their approach with a phone surveys of a geographically stratified random sample of 856 individual subscribers and using a DHS composite wealth index at micro-region level. In \citet{pokhriyal2017combining}, we explore this topic further.

\subsection{Transportation}

Using CDR data from the first D4D for the city of Abidjan (Ivory Coast), \citet{berlingerio2013allaboard} evaluated which new routes would best improve the existing transit network to increase ridership and user satisfaction, both in terms of reduced travel and wait time. Four new routes have been proposed by the optimization system (called AllAboard), resulting in an expected reduction of 10\% city-wide travel times.

\subsection{Energy}

The first prize of the D4D Senegal challenge was awarded to a research project which assessed the contribution of mobile phone data for the development of bottom-up energy demand models in Senegal \citep{martinez2015using}. Specifically, the research team introduced a framework that combines mobile phone data analysis (mobile phone activity was used as a proxy of the energy consumption), socio-economic, geo-referenced data analysis, and state-of-the-art energy infrastructure engineering techniques to assess the techno-economic feasibility of different centralized and decentralized electrification options for rural areas in a developing country. The result was a country map of electrification recommended option between (i) extensions of the existing medium voltage grid, (ii) diesel engine-based community-level Microgrids, and (iii) individual household-level solar photovoltaic systems.







\section{Statistical Limitations}

CDRs are a good example of Big Data source that can be diverted from their primary purpose to approximate socio-economic variables and population mobility. As they are not designed for this purpose, this means that an unavoidable bias will always impact any application based on these data. If not properly understood, this could lead to serious misinterpretation of the results and ultimately, have harmful impacts in misleading policy-makers. This section reviews some of the sources of inaccuracy inherent to mobile phone metadata.

\subsection{Technical issues}  

MNOs suffer occasional down-time during which data are not recorded (missing data). Furthermore, cells can also be deactivated for maintenance or resource optimization (e.g., during low activity period such as nighttime). On the other hand, incorrect data can arise at different level of the data collection due to encoding or other technical issues (e.g. duplicated records, records with incorrect time values, etc.).

\subsection{Selection bias} 

\begin{figure*}[t!]
    \centering        
        \includegraphics[width=0.8\linewidth]{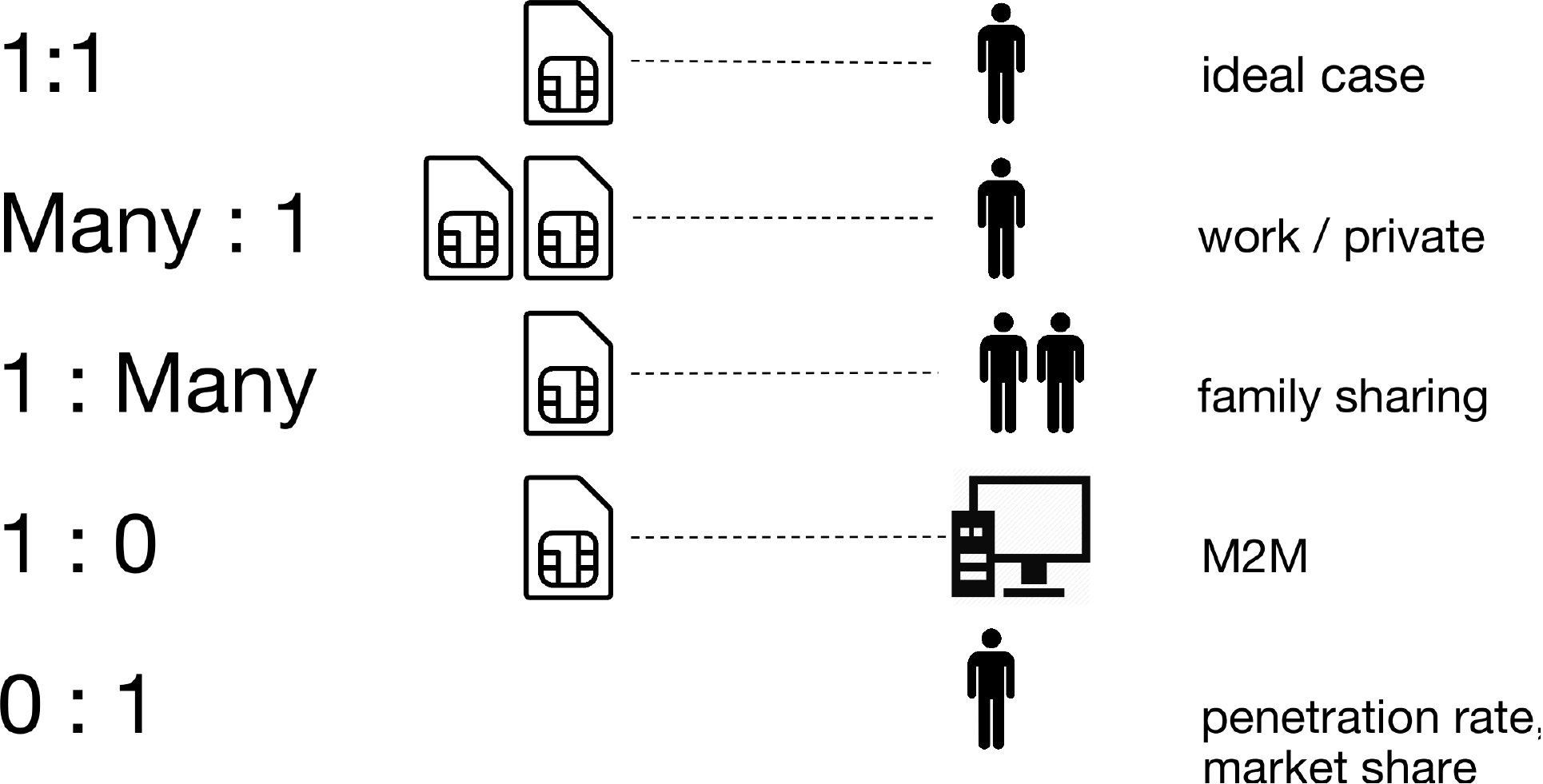} \\ 
    \caption{Possible association schemes between SIM and persons. Figure adapted from \citet{ricciato2015estimating}. M2M stands for Machine to Machine communication. }
    \label{fig:SIMsAndPersons}
\end{figure*}

People generating CDR data have selected themselves as data generators through their activity. This is called a ‘selection bias’. First, while the penetration of mobile phone is very high in the developing world, some sociodemographic groups (typically young
children and senior people) are still left out of the analysis when considering mobile phone metadata. The adoption base in Africa has been more traditionally skewed towards a wealthier, educated, urban and predominantly male population \citep{aker2010mobile,blumenstock2010mobile}. Additionally, one SIM card does not necessarily correspond to one person. Figure \ref{fig:SIMsAndPersons} illustrates all the possible association schemes between SIM and persons. In developing countries, this is frequent that someone owns different SIM cards to be able to switch between mobile carrier's network depending on promotional campaigns. Phone sharing is also a common practice among the poorest. On the other hand, data access is most often limited to only one provider in the country. This could be problematic if the choice of a MNO is correlated with the socio-economic profile of individuals.  \\

However, depending on the application, the impact of ownership bias might not be as strong as expected. For instance, \citet{wesolowski2013impact} show that mobility estimates are surprisingly robust to the substantial biases in phone ownership across different geographical and socio-economic groups using 1-year data of 15 million individuals in Kenya.

\subsection{Spatial bias}

Lacking data on antenna power and orientation, their coverages are generally approximated by means of a Voronoï tessellation (Figure \ref{fig:networkCDR} -- A). It assumes that mobile phones always connect to the closest antenna. However, dozens of factors play a role in the decision of the system to assign a specific cell to a phone (e.g., signal strength, atmospheric conditions, traffic overload, maintenance schedule). It is, therefore, possible that a person at the same location, making five successive phone calls, will connect to five different antennas. Designed for business and not tracking, CDRs provide information that helps companies manage their operations, not track phones. On the other hand, as it has been already mentioned, the spatial resolution of CDRs depends on the BTS density. It means that remote and unpopulated areas, where populations at risk (such as poor and food insecure population) are generally found, have lower spatial resolution than urban areas due to a lower antenna density. Finally, CDR data are always limited to one country and due to technical challenge, the cross-border movements are difficult to capture. This is an important limitation for large scale epidemiological studies because the spread of a disease does not stop at the border. Other similar data sources capturing geographic digital footprints (e.g. tweets) may be used to overcome this limitation \citep{blanford2015geo}.

\section{Data Access}

Because private companies hold the data, there is no guarantee of access. It requires an agreement between researchers and the telecom company. The companies might be reluctant to provide access due the threat to subscribers privacy that can result in a loss of customers. Therefore, their interest to open/sell datasets is somewhat limited. Yet, it is generally possible to get access to a particular data set for testing/research purposes, but still far harder (for legal or commercial reasons) to extend this access for production purposes. \\

Research teams and institutions have learned the hard way that even in case of emergency situations, being granted access to CDRs can still remain an unsurmountable obstacle. When the Ebola epidemic broke out in Sierra Leone, Liberia, and Guinea in 2014, a group of academics and international development actors began to call for the use of aggregated location data from mobile phone networks in order to facilitate the response effort \citep{Economist2014}. After dozens of conference calls over many months involving over fifty participating organizations (including several UN agencies), permission was finally granted by the relevant local authorities -- except for Liberia. Despite having the highest death toll of any country that experienced the Ebola epidemic, Liberia never released CDRs, in part due to concern about their ability to enforce privacy protection \citep{mcdonald2016ebola}. This experience raises fundamental questions on the trade-off between privacy and utility and how it can be adjusted according to the level of emergency of a situation. \\

In the aim to respond to these concerns, the mobile phone industry association -- the GSMA -- has developed a ‘Mobile for Development Intelligence’ programme to persuade mobile providers from developing countries to share data with researchers, industries and development organisations. However, their focus is primarily on commercial outputs as the goal of their open data portal -- Mobile for Development Products and Services Trackers --  is worded as:

\begin{displayquote}
“offer the industry access to high quality data to help improve business decision making, increase total investment from both the commercial mobile industry and the development sector as well as to accelerate economic, environmental and social impact from mobile solutions.“ \citep{Metcalfe2015}
\end{displayquote}

\section{Data Privacy}

A lot of personal information can be extracted from CDRs. Using such data, it is easy to know where people live and work as well as tracking most of their movement \citep{calabrese2013understanding, ahas2010using}. Their social network can be characterized allowing, for instance, the examination of the evolution of relationships over time \citep{eagle2009inferring}. The way people use their phone is also a good indicator of their personality. For instance, \citet{de2013predicting} showed how CDRs could be used to infer five main traits of personality: openness, conscientiousness, extraversion, agreeableness, and neuroticism (a socio-psychological model known as OCEAN). Based on facebook data, the same model was used by Cambridge Analytica to micro-target of campaign material to US voters with the purpose of influencing the 2016 presidential campaign \citep{youyou2015computer, grassegger2017data}.\\

To protect people's privacy, mobile phone data are always anonymized, i.e, all personal data such as name, address, phone number, etc., are either removed from the database or replaced by a randomly generated number to avoid identification\footnote{This process is known as pseudo-anonymization.}. Data are then provided to a third party after a non-disclosure agreement (NDA) was signed with the MNO. The purpose of the agreement is to prevent CDRs to be shared to another party, and to define the scope of research questions that will be explored with the data. Both the anonymization procedure and the NDA are supposed to preserve the safety of users privacy. \\

However, if individual patterns are unique enough, additional information can be used to link the data back to an individual. 
Using fifteen months of human mobility data derived from CDRs, \cite{de2013unique} showed that four randomly chosen points (i.e., four places where a user was at a specific time) are enough to uniquely characterize 95\% of the users, whereas two randomly chosen points still uniquely characterize more than half of the users. \\

\begin{figure*}[t!]
    \centering        
        \includegraphics[width=0.3\linewidth]{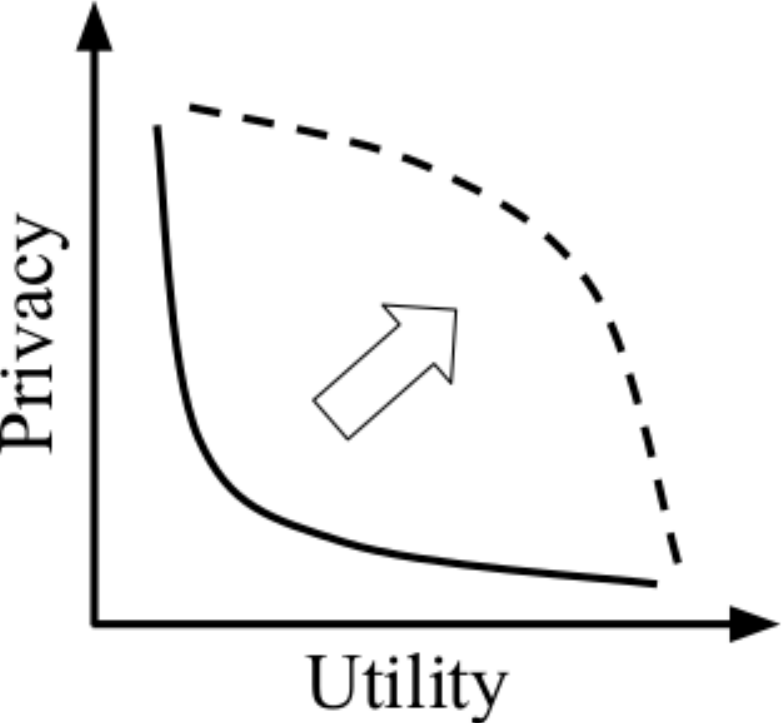} \\ 
    \caption{Schematic representation of the trade-off between privacy and utility of personal data. Full line is the actual relationship and the dotted line shows the ideal relationship. The figure is adapted from an OPAL presentation (\url{www.opalproject.org}).}
    \label{fig:utility}
\end{figure*}

Data aggregation allows to further strengthen privacy. In the case of CDRs, several approach can be used. First, users can be aggregated by BTS. With this aggregation, it is no more possible to track one specific user and mobility analyses cannot be performed anymore. On the other hand, it is still relevant for dynamic population mapping as it only requires the number of a users at a specific place and time. This is also still useful to study spatial network based on antenna-to-antenna traffic. To keep mobility analysis feasible, temporal and/or spatial aggregation can be used. However, decreasing the resolution comes with a loss in data utility so that a trade-off exists between privacy protection and the preservation of data value (Figure \ref{fig:utility}). Furthermore, \cite{de2013unique} showed that blurring the spatial and temporal resolution does not significantly impact the number of points needed to re-identify a user in the database. Finally, to preserve privacy, noise can also be added to some variables of the database (e.g. random spatial reallocation of BTS).

On the other hand, in today's digital world, increasing privacy is only useful if it is done for all sources at the same time. This issue was defined as the ’secondhand smoke problem’ by \citet{lane2010survey}: 

\begin{displayquote}
“[...] the secondhand smoke problem of mobile sensing creates new privacy challenges, such as: 

• How can the privacy of third parties be effectively protected when other people
wearing sensors are nearby?

• How can mismatched privacy policies be managed when two different people are close enough to each other for their sensors to collect information from the other party?“
\end{displayquote}

With the aim to solve some of these issues, the Open Algorithm (OPAL) project was set up by Orange, MIT Media Lab, Data-Pop Alliance, the World Economic Forum and the Imperial College London \citep{hardjono2016trust}. The idea behind OPAL is to bring about a paradigm shift in mobile metadata analysis by moving the code to the data rather than data to the code. It means that instead of providing data directly to researchers through NDA, CDRs (or other sensitive data) remain behind the firewall of each provider. Only certified algorithms, meeting predetermined privacy standards, can be run on the data in this secure environment and only aggregated results are shared with the user (\url{www.opalproject.org}). It allows facilitating data access while preserving business and individual privacy. In \citet{pokhriyal2017combining}, we illustrated a similar approach where two models of poverty mapping based on disparate data sources can be combined without the need to share the raw data.

\section{Conclusions}

In this paper, the specific features of mobile phone metadata were discussed with a focus on applications for development. The amount of information held in these data is fantastic. Among other, they have been used to model the spread of infectious diseases, study road traffic, support electrification planning strategies or map the socio-economic level of population. While massive, CDRs are not statistically representative of the whole population due to several sources of bias. Furthermore, data access and privacy are significant challenges that are not necessarily straightforward to resolve.\\

While the challenges exist, the potential of such data might exceed the limitations. Compared to traditional data collected to compute official statistics, they are cost-effective and can provide faster or even near real-time insights. They might also be used to test concepts and define future research questions. On the other hand, the Reality Mining project demonstrated that observed behavior using mobile phone metadata strongly differs from what was self-reported by the same individuals \citep{eagle2006reality}. This suggests that the subjectivity of the subjects’ perception produces a significant bias in traditional surveys. The objectivity coming from their exogeneity is, therefore, another strength of mobile phone metadata.

\newpage
\section*{References}
\bibliographystyle{model1-num-names}
\bibliography{sample.bib}







\end{document}